\documentclass[11pt]{article}
\usepackage[a4paper]{geometry}

\usepackage{amsmath}
\usepackage{amssymb}
\usepackage{amsthm}
\usepackage{mathrsfs}
\usepackage{bbm}
\usepackage{empheq}
\usepackage[loose]{subfigure}
\usepackage{epsfig}
\usepackage{graphicx}
\usepackage{psfrag}
\usepackage[usenames,dvipsnames]{pstricks}
\usepackage{pst-plot}
\usepackage[colorlinks]{hyperref}
\usepackage{tabls}
\usepackage{paralist}
\usepackage{hyperref}
\usepackage{comment}

\DeclareSymbolFontAlphabet{\Bbb}{AMSb}

\newlength{\fixboxwidth}
\newcommand{\COMMENT}[1]{}

\renewcommand{\P}{\mathbb{P}}

\newcommand{\quark}{\setbox0\hbox{$x$}\hbox to\wd0{\hss$\cdot$\hss}}

\newcommand{\pa}{\text{particle}}
\newcommand{\wa}{\text{wave}}

\newtheorem{thm}{Theorem}

\theoremstyle{definition}

\newtheorem{rmk}[thm]{Remark}

\newtheorem{Strategy}{Strategy}

\hypersetup{
    linkcolor=blue,
}

\title{On testing the simulation theory}

\author{Tom Campbell\footnote{twcjr44@gmail.com}, Houman Owhadi\footnote{California Institute of Technology, Computing \& Mathematical Sciences , MC 9-94 Pasadena, CA 91125, owhadi@caltech.edu}, Joe Sauvageau\footnote{Jet Propulsion Laboratory, California Institute of Technology, 4800 Oak Grove Drive, Pasadena, California 91109, jsauvage@jpl.nasa.gov. This work was done as a private venture and not in the author's capacity as an employee of the Jet Propulsion Laboratory, California Institute of Technology.}, David Watkinson\footnote{Main Street Multimedia, Inc. 3005 Main St. \#406, Santa Monica, CA 90405, watkinsondavidm@gmail.com}}

\date{\today}

\renewcommand{\thefigure}{\arabic{figure}}

\makeatletter
\renewcommand{\p@subfigure}{\thefigure}
\makeatother

\newcounter{mycount}

\makeatletter
\def\blfootnote{\gdef\@thefnmark{}\@footnotetext}
\makeatother

\begin{document}

\maketitle

\begin{abstract}
Can the theory that reality is a simulation   be tested? We investigate this question based on the assumption that if the system performing  the simulation is finite (i.e. has limited resources), then to achieve low computational complexity, such a system would, as in a video game, render  content (reality) only at the moment  that information becomes available for observation
by a player and not at the moment of detection by a machine (that would be part of the simulation and whose \emph{detection} would also be part of the internal computation performed  by the Virtual Reality server before rendering content to the player). Guided by this principle we describe  conceptual wave/particle duality experiments aimed at testing the  simulation theory.
\end{abstract}

\section{Introduction}
 Wheeler advocated \cite{Wheeler1990} that ``Quantum Physics requires a new view of reality'' integrating physics with  digital (quanta) information. Two such views emerge from the presupposition that reality could be computed. The first one, which includes
Digital Physics \cite{Zuse1967} and the cellular automaton interpretation of Quantum Mechanics  \cite{HooftG2016}, proposes that the universe \emph{is} the computer. The second one, which includes the simulation theory \cite{BOSTROM2003, Campbell2007, WhitworthB2007} (also known as the simulation hypothesis), suggests that the observable reality is entirely virtual and the system performing the simulation (the computer) is distinct from its simulation (the universe). In this paper we investigate the possibility of experimentally testing the second view, and base our analysis on the assumption that the system performing the simulation has limited computational resources. Such a system would therefore use computational complexity as a minimization/selection principle for algorithm design.

Although, from a broader perspective, the issue being addressed is the nature of reality, to avoid philosophical positions, the subject of this paper and of our tests will be limited to the following question:
\begin{quote}
What causes and determines the collapse of the wave function?
\end{quote}

\section{Review}
There are three main perspectives on the nature and causation of the collapse of the wave function.
\begin{enumerate}
\item The Copenhagen interpretation  \cite{schlosshauer2013snapshot}
 states that physicists can calculate as if quantum waves existed, even though they do not. In this view, any detection causes a quantum collapse that does not actually occur.
\item In the many worlds theory \cite{everett1957relative, dewitt1973many}, which views reality as a branching process where measurements cause splitting into possible outcomes,
 there is no quantum collapse because every collapse option occurs in some other universe.

\item In the	Von Neumann-Wigner theory \cite{von1955mathematical, wigner1967symmetries} human consciousness triggers quantum collapse.

\end{enumerate}

The Copenhagen interpretation, summed up by D. Mermin \cite{mermin1989} as ``Shut up and calculate!'', renounces addressing the question. The many worlds theory is incredibly inefficient from a computational complexity perspective.
The theory that consciousness causes collapse is not widely accepted as it seems to lead towards solipsism  and problems of consistency over time (which caused Wigner to abandon \cite{esfeld1999essay} this theory towards the end of his life):
why create an event (e.g. the big-bang) with no-one there to observe it?

 \paragraph{On QBism.}
A theory that is  related to the simulation theory emerges from the interpretation of the wave function as a  description of a single observer’s subjective knowledge about the state of the universe and the interpretation of the wave function as a process of conditioning this subjective prior. This theory,  known as QBism \cite{mermin2014physics},
 is  the  Bayesian formulation of Quantum Mechanics, and it is
 distinct from the Copenhagen interpretation  \cite{mermin2017qbism} and the 	Von Neumann-Wigner interpretation in the sense that the
  ``wave function is not viewed as a description of an objective reality shared by multiple observers but as a description of a single observer’s subjective knowledge'' \cite{qbismprivate}. As in the simulation theory,
  QBism  resolves quantum paradoxes at the cost of realism, i.e. by embracing the idea that a single objective reality is an illusion.

\section{Theory description }
We will now describe the simulation theory from a computational perspective. Although we may call the system performing the computation \emph{computer}, \emph{VR server/engine}, or \emph{Larger Consciousness}, the specification of the fundamental nature of this system is not necessary to the description of the theory. Our core assumption is that this system is finite and, as a consequence, the content of the simulation is limited  by (and only by) its finite processing resources and the system seeks to achieve low computational complexity.

\paragraph{On rendering reality}
It is now well understood in the emerging science of Uncertainty Quantification \cite{handbookofuq2017} that low complexity computation
must be performed with hierarchies of multi-fidelity models \cite{multifidmod2016}.
It is also now well understood, in the domain of game development,  that low computational complexity requires rendering/displaying content only when observed by a player. Recent games, such as \emph{No-Man's Sky} and \emph{Boundless}, have shown that vast open universes (potentially including ``over 18 quintillion planets with their own sets of flora and fauna'' \cite{Ravicnn2015}) are made feasable by creating content, only at the moment the corresponding information becomes available for observation
 by a player, through randomized generation techniques (such as procedural generation).
Therefore, to minimize computational complexity in the simulation theory,  the system performing the simulation would render reality only at the moment the corresponding information becomes available for observation  by a conscious observer (a player), and the resolution/granularity of the rendering would be adjusted to the level of perception of the observer. More precisely, using such techniques, the complexity of simulation would not be constrained by  the apparent size of the universe or an underlying pre-determined  mesh/grid size  \cite{Beaneal2014} but by the number of players and the resolution of the information made available for observation.

\paragraph{On the compatibility of the simulation theory with Bell's \emph{no go theorem}.}
Bell's \emph{no-go theorem} shows that the predictions of quantum mechanics cannot be recovered/interpreted, in terms of classical probability through the introduction of \emph{local} random variables. Here, the ``\emph{vital assumption}'' \cite[p.~2]{Bell1964a} made by Bell is the absence of action at distance (i.e. as emphasized in \cite[eq.~1]{Bell1964a}, the independence of the outcome of an experiment performed on one particle, from the setting of the experiment performed on another particle).
Therefore Bell's no-go theorem does not prevent a (classical) probabilistic interpretation of quantum mechanics using a  ``spooky action at distance'' \cite{EPR1935}.
Here, the simulation theory offers a very simple explanation for the violation of the principle of locality implied by Bell's \emph{no-go theorem} \cite{Bell1964a}, the EPR paradox \cite{EPR1935}, Bell's inequalities violation experiments \cite{aspect1982experimental, aspect1982experimentalbis} and quantum entanglement \cite{horodecki2009quantum}: notions of locality and distance defined within the simulation do not constrain  the action space of the system performing the simulation (i.e. from the perspective of the system performing the simulation, changing the values of variables of spins/particles separated by 1 meter or 1 light year has the same complexity).

\paragraph{On the emergence of probabilistic computation.}
It is well understood in Information Based Complexity (IBC) \cite{Traub1988, Woniakowski1986, Packel1987, Nemirovsky1992, Woniakowski2009} that low complexity computation requires computation with partial/incomplete information.
As suggested in \cite{Packel1987} and shown in \cite{OwhadiMultigrid:2015} the identification of near optimal complexity algorithms
requires playing repeated adversarial (minimax) games against the missing information. As in Game  \cite{VNeumann28, VonNeumann:1944} and Decision Theory \cite{Wald:1945}, optimal strategies for such games are randomized strategies \cite{OwhadiMultigrid:2015}. Therefore Bayesian computation emerges naturally \cite{OwSccig2017, cockayne2017bayesian} in the presence of incomplete information (we refer to
 \cite{Poincare:1896, Suldin1959, Larkin1972, Diaconis:1988, Shaw:1988, Hagan:1991, Owhadi:2014, Hennig2015, OwScWald2016, OwhadiMultigrid:2015, cockayne2017bayesian, OwSccig2017} for a history of the correspondence between Bayesian/statistical inference, numerical analysis and algorithm design).
Given these observations the fact that quantum mechanics can naturally be interpreted as Bayesian analysis with complex numbers
 \cite{Caves2002-CAVQPA, Benavolietal2016} suggests its natural interpretation as an optimal form of computation in presence of incomplete information.
 Summarizing, in the simulation theory,
 to achieve near optimal computational complexity by computing with partial information and limited resources, the system performing the simulation would have to \emph{play dice}.
It is interesting to note that in \cite{Benavolietal2016} the Bayesian formulation of Quantum Mechanics is also logically derived in a game theoretic setting.

 \paragraph{On the initialization of the simulation and solipsism.}
 A universe simulated by a finite system would have a necessary beginning (e.g., a big bang or pressing the \emph{Enter} key) that cannot be explained from a perspective confined to the simulation itself.
 In this theory, the consciousness of the players form the screen on which reality is rendered and problems of
  solipsism are resolved from the perspective  of a multi-player VR game: if a tree falls in a forest when no player is there to observe it,  its fall is only part of the internal computation performed  by the Virtual Reality server before rendering content to the player
and fine details (mold, fungi, termites, a rat’s nest in it with babies) are only rendered at that moment to preserve the consistency of the simulation (assuming that the rats are non-playing characters).

\paragraph{What causes and determines the collapse of the wave function?}  Or in Virtual Reality (VR) terminology, what causes the virtual reality engine to compute and make information defining the VR available to an experimenter within the VR?
 Is it
\begin{enumerate}[(I)]
\item entirely determined by the experimental/detection set-up?
\item or does the observer play a critical role in the outcome?
\end{enumerate}
In the simulation theory, these questions can be analyzed based on the idea that a good/effective VR would operate based on two, possibly conflicting, requirements:
 (1) preserving the consistency  of the VR (2) avoiding detection (from the players that they are in a VR).
  However, the resolution of such a conflict would be limited by computational
resources, bounds on computational complexity, the granularity of the VR being rendered and logical constraints on how inconsistencies can be resolved.  Occasionally, conflicts that were unresolvable would lead to VR indicators and discontinuities (such as the wave/particle duality).

Summing up, to save itself computing work, the system only calculates reality when information  becomes available for observation by a player, and to avoid detection by players it maintains a consistent world, but occasionally,
 conflicts that are unresolvable  lead to VR indicators and discontinuities (such as the wave/particle duality).
  Although this perspective supports the
  	Von Neumann-Wigner postulate that \cite{von1955mathematical, wigner1967symmetries} human consciousness is necessary for the completion of quantum theory, the simulation theory also agrees with Copenhagen in the sense that it does not require the actual existence of quantum waves or their collapse (which are seen as useful predictive models made by the players immersed in the VR).

\section{Hypothesis test}

Two strategies can be followed to test the simulation theory: (1) Test the moment of rendering (2) Exploit conflicting requirement of  logical consistency preservation and  detection avoidance to force the VR rendering engine  to create discontinuities in its rendering or  produce a measurable signature event within our reality that indicates that our reality must be simulated.

\begin{figure}[h!]
	\begin{center}
			\includegraphics[width=0.75\textwidth]{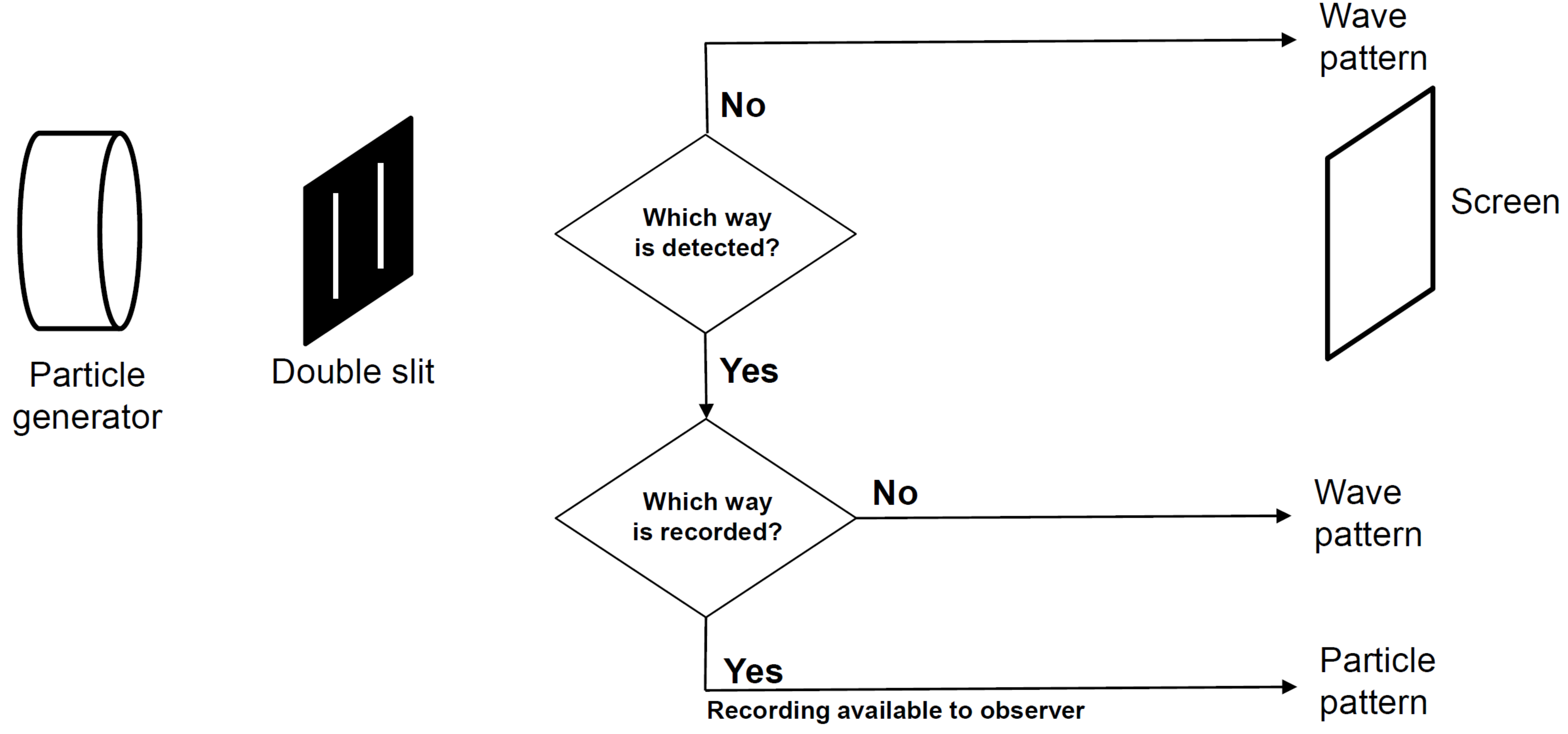}
		\caption{Detecting but not recording ``which-way''}\label{figrecordoff}
	\end{center}
\end{figure}

\paragraph{Testing the moment of rendering.}
In subsections  \ref{secpredict2}, \ref{secpredict3} and \ref{secpredict} we will describe wave-particle duality experiments (illustrated in figures  \ref{figrecordoff}, \ref{figsub} and \ref{figsuperdelayedchoicequantumeraser})  aimed at testing the
 simulation theory by testing the hypothesis that reality is not rendered (or the wave function is not collapsed) at the moment of detection by an apparatus that would be part of the simulation, but rather at the moment when the corresponding information becomes available for observation by an experimenter. More precisely, in the setting of wave-particle duality experiments, our hypothesis is that wave or particle duality patterns are not determined at the moment of detection but by the existence and availability of the \emph{which-way} data when the pattern is observed.

\paragraph{Exploiting consistency vs detection.}
In Subsection \ref{seclkjhjhu} we propose though experiments  where the conflicting requirement of  logical consistency preservation and  detection avoidance is exploited to force the VR rendering engine  to create discontinuities in its rendering or  produce a clear and measurable signature event within our reality that would be an unambiguous indicator that our reality must be simulated. Although we cannot predict the outcome of the experiments proposed in Subsection \ref{seclkjhjhu} we can rigorously prove that their outcome will be new in comparison to classical wave-duality experiments.
As a secondary purpose, the analysis of the experiment of Subsection \ref{seclkjhjhu} will also be  used
 clarify the notion of availability of \emph{which-way} data in a VR.

\begin{figure}[h!]
	\begin{center}
			\includegraphics[width=0.75\textwidth]{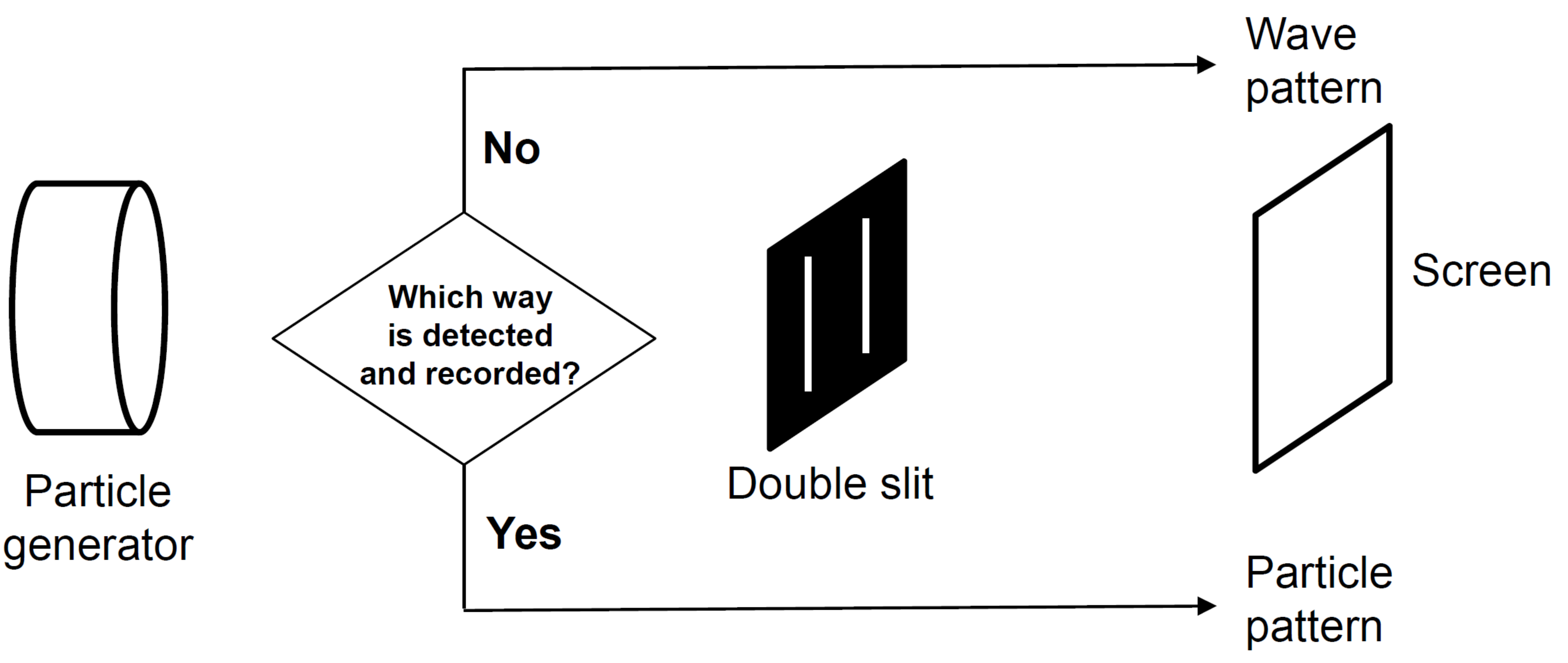}
		\caption{The classical double slit experiment \cite{feynmanvol3, aspect1987wave} with \emph{which-way} detected  before or at the slits. We write ``wave pattern'' for interference patten, and ``particle pattern'' for non-interference pattern. }\label{figdoubleslit}
	\end{center}
\end{figure}
 \begin{figure}[h!]
	\begin{center}
			\includegraphics[width=0.75\textwidth]{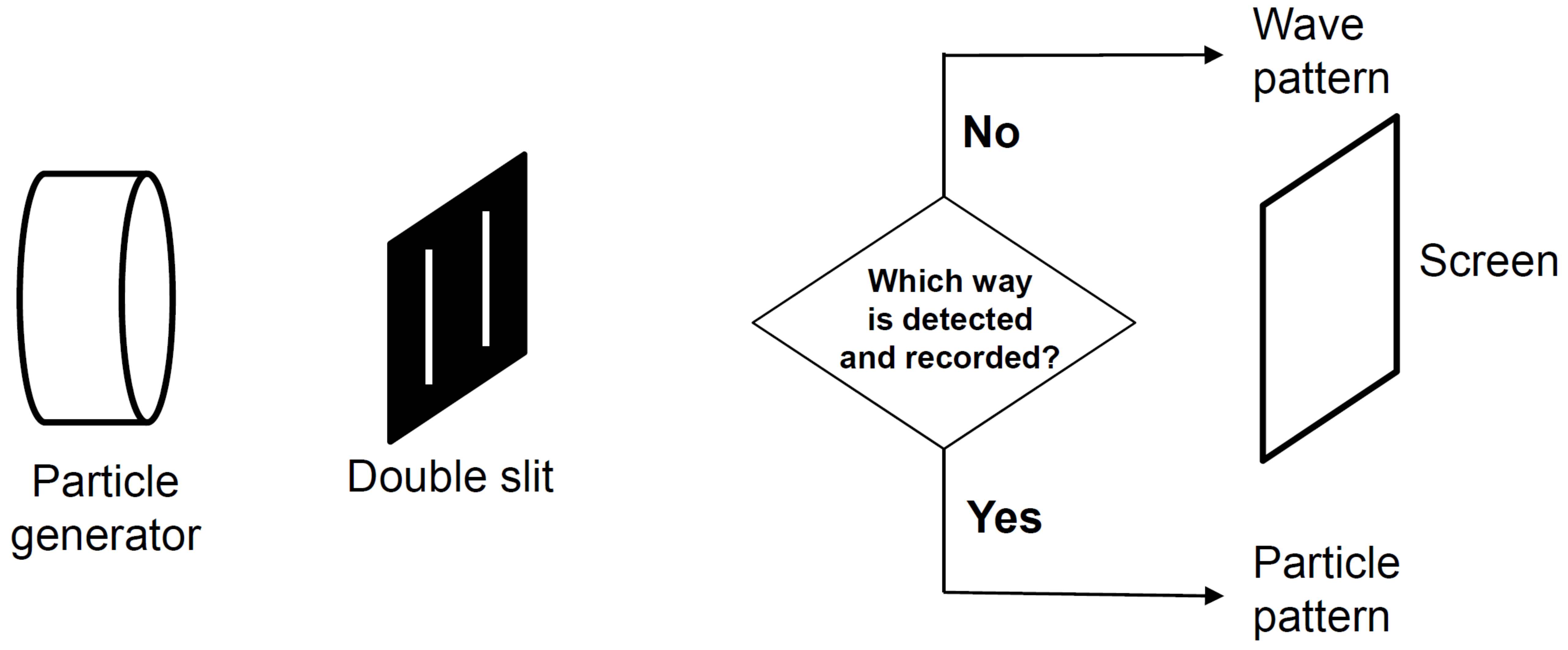}
		\caption{The delayed choice experiment \cite{wheeler1978past, jacques2007experimental}. The choice of whether or not to detect and record \emph{which-way} data is delayed until after each particle has passed through a slit but before it reaches the screen.}\label{figdelayedchoice}
	\end{center}
\end{figure}
 \begin{figure}[h!]
	\begin{center}
			\includegraphics[width=0.75\textwidth]{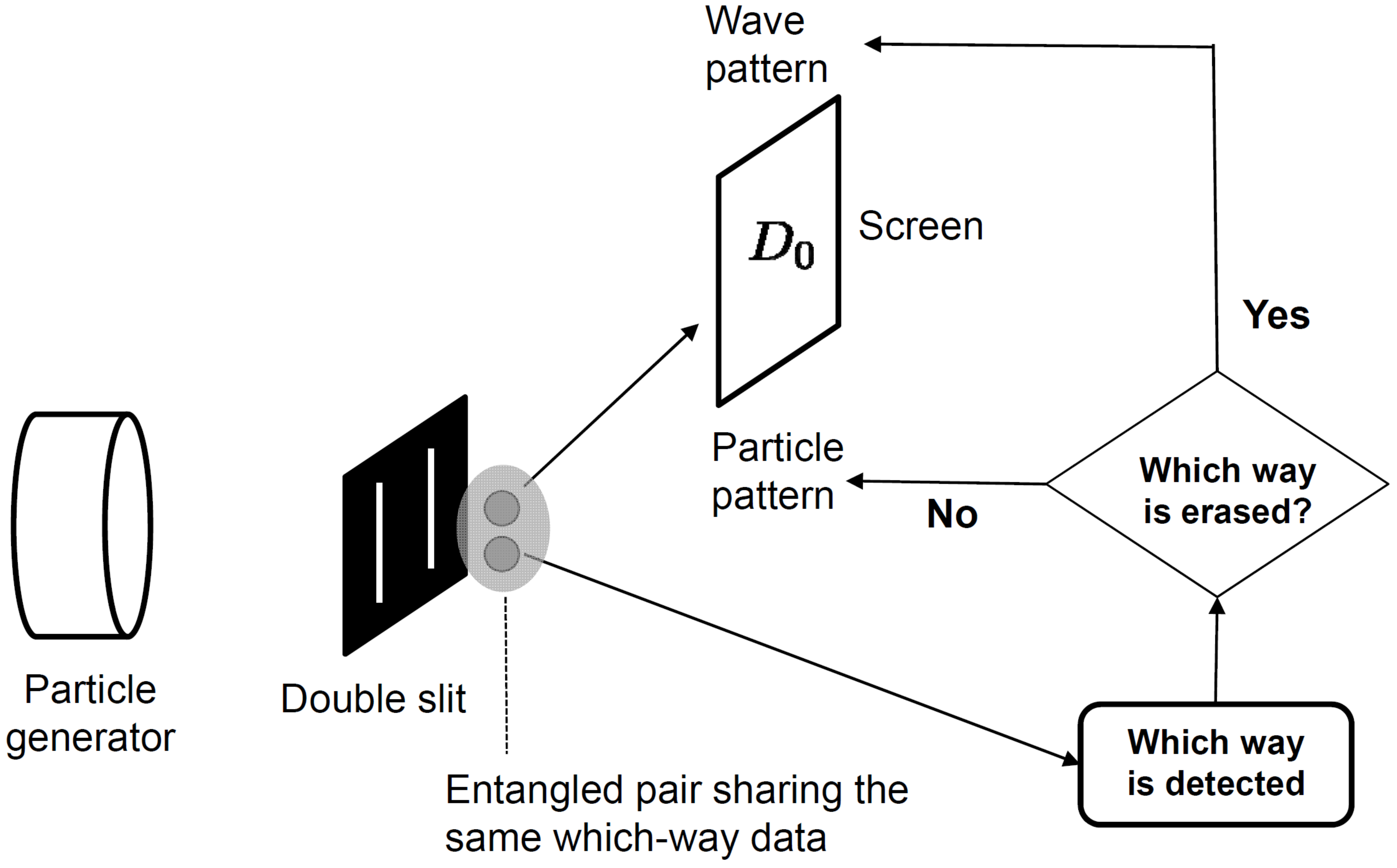}
		\caption{The delayed choice quantum eraser experiment \cite{scully1982quantum, kim2000delayed}. \emph{Which-way} data is collected before, at, or after each particle has passed through a slit, however, this \emph{which-way} data may be erased before the particle hits the screen.   This experiment is sometimes called a delayed erasure experiment since the decision to erase is made after the particle has passed through a slit (chosen one path or the other).}\label{figdelayedchoicequantumeraser}
	\end{center}
\end{figure}

\subsection{Wave-particle duality experiments}
Although the double slit experiment has been known as a classic thought experiment \cite{feynmanvol3} since the beginning of quantum mechanics, and although this experiment was performed with ``feeble light'' \cite{taylor1909interference} in 1909 and electron beams \cite{jonsson1961elektroneninterferenzen} in 1961, the first  experiment with single photons was not conducted prior to 1985 (we refer to \cite{aspect1987wave}, which also describes how the interpretation of ``feeble light'' experiments in terms of quantum mechanics is ambiguous due to the nature of the Poisson distribution associated with ``feeble light'').
The double slit experiment, illustrated in its simplified and conceptual form in Figure~\ref{figdoubleslit}, is known as the classical demonstration of the concept of wave/particle  duality of quantum mechanics. In this classical form, if \emph{which-way} (i.e.   which slit does each particle pass through) is   ``detected and recorded'' (at the slits), then  particles (e.g. photons or electrons) behave like particles and a non-diffraction pattern is observed on the screen. However, if \emph{which-way} is not  ``detected and recorded,'' then  particles behave like waves and an interference pattern is observed on the screen.
Since in the classical set up, the \emph{which-way} detection is done at the slits, one may wonder whether the detection apparatus itself, could have induced the particle behavior, through a perturbation caused by its interaction with the photon/electron going through those slits.  Motivated by this question, Wheeler  \cite{wheeler1978past} argued, using a thought experiment (illustrated in its simplified and conceptual form in Figure \ref{figdelayedchoice}), that the   choice to perform the \emph{which-way} detection could be delayed and done after  the double-slits. We refer to \cite{jacques2007experimental} for the experimental realization of Wheeler's delayed-choice gedanken experiment. Comparing Figure \ref{figdelayedchoice} with Figure \ref{figdoubleslit}, it appears that whether the \emph{which-way} data is detected and recorded before, at, or after the slits makes no difference at the result screen.
In other words, the result at the screen appears to not be determined by when or how that \emph{which-way} data is detected but by
having the recorded \emph{which-way} data before a particle impacts that screen.

 Following Wheeler, Scully and Dr\"{u}hl \cite{scully1982quantum} proposed and analyzed an experiment (see Figure \ref{figdelayedchoicequantumeraser}), realized in \cite{kim2000delayed}, where the \emph{which-way} detection is always performed ``after the beam has been split by appropriate optics'' and the screen data has been collected,  but before it is possibly erased (with probability $1/2$ using a beam-splitter).  We also refer to \cite{ma2013quantum} for a set-up with significant separation in space between the different elements of the experiment.
Comparing Figure \ref{figdelayedchoicequantumeraser} with Figure \ref{figdelayedchoice}, it appears that whether the \emph{which-way} data is or is not erased determines the screen result. Again, the result at the screen seems to be determined, not by the detection process itself but by the availability of  the \emph{which-way} data.  Erasing the \emph{which-way} data appears to be equivalent to having never detected it.

 \begin{figure}[h!]
	\begin{center}
			\includegraphics[width=0.75\textwidth]{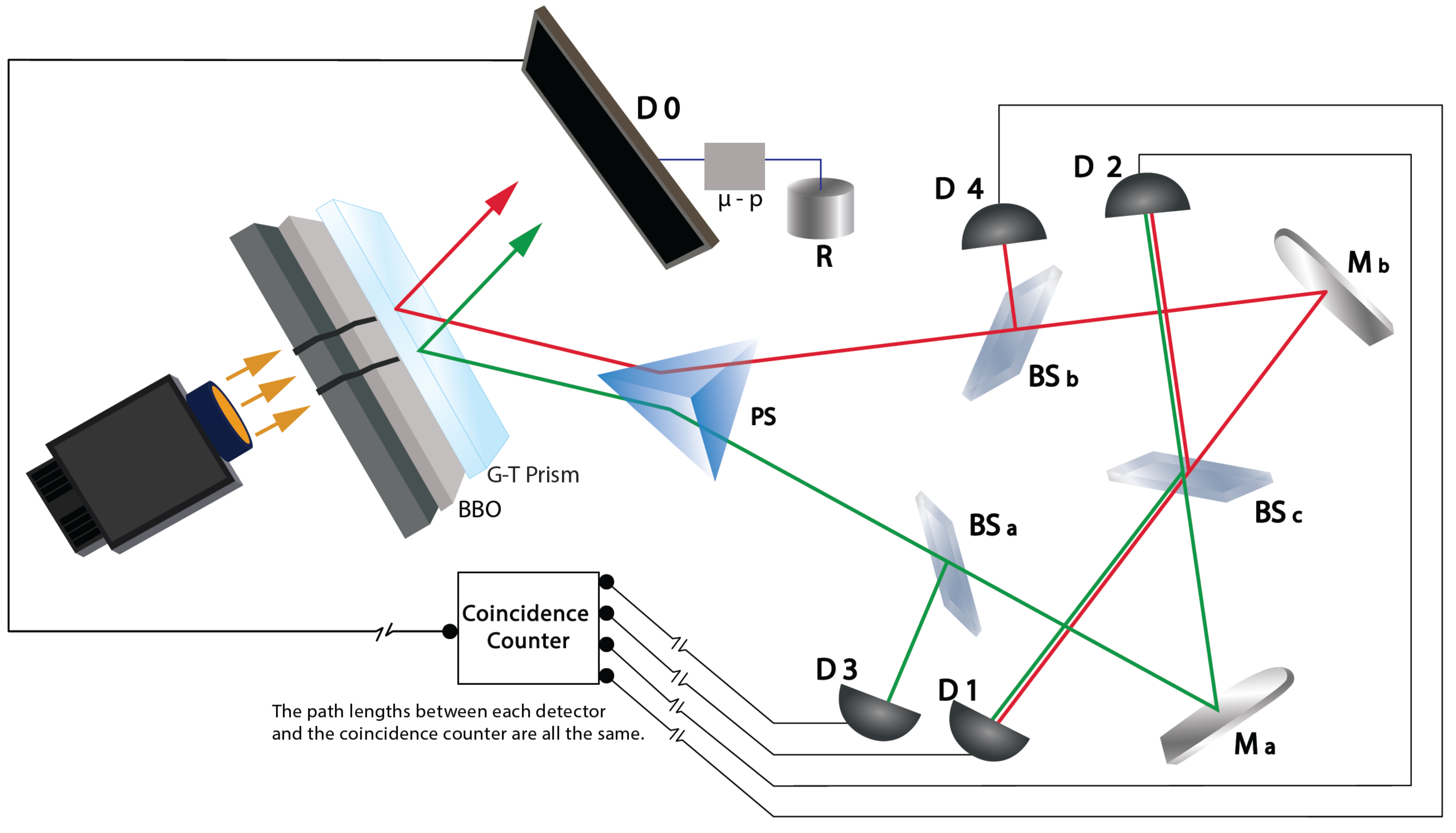}
		\caption{The delayed choice quantum eraser experiment set up as described in \cite{kim2000delayed}. The  microprocessor ($\mu$-p) represents an addition to the original experiment and will be discussed in subsections \ref{secpredict} and  \ref{seclkjhjhu}.
}\label{figerasuresetup}
	\end{center}
\end{figure}
\begin{rmk} \label{rmkfeature}
A  remarkable feature of the delayed choice quantum eraser experiment \cite{kim2000delayed} (see figures \ref{figdelayedchoicequantumeraser} and \ref{figerasuresetup}) is the creation of an entangled photon pair  (using a type-II phase matching nonlinear optical crystal BBO: $\beta - BaB_2O_4$) sharing the same \emph{which-way} data and the same creation time. One photon is used to trigger the coincidence counter (its impact location screen $D_0$ is also recorded) and the second one is used to detect the \emph{which-way} data and possibly erase it (by recording its impact on  detectors $D_1,D_2, D_3$ and $D_4$). The coincidence counter is used to identify each pair of entangled photon by tagging each impact on the result screen $D_0$ and each
event on the detectors $D_1,D_2, D_3$ and $D_4$ with a time label.
Using the coincidence counter to sort/subset the impact locations (data) collected on the result screen $D_0$, by
 the name ($D_1,D_2, D_3$ or $D_4$) of the detector activated by the entangled photon, one obtains the following patterns:

\begin{enumerate}[D1:]
\item  Interference pattern (\emph{which-way}  is erased).
\item Interference pattern (\emph{which-way}   is erased).
\item  Particle pattern (\emph{which-way}   is known, these photons are generated at Slit 1).
\item  Particle pattern (\emph{which-way}   is known, these photons generated at Slit 2).
\end{enumerate}
\end{rmk}

\subsection{Detecting but not making the data available to  an observer}\label{secpredict2}
The following experiments are designed based on the hypothesis that the availability of  \emph{which-way} data to an observer is the key element that determines the pattern found on the result screen: the simulated content (the virtual reality) is computed and available to be rendered to an \emph{experimenter}
only at the moment that information becomes available for observation by an \emph{experimenter} and not at the moment of detection by an apparatus.

In the proposed experiment, illustrated in a simplified and conceptual form in Figure \ref{figrecordoff}, the \emph{which-way} data is detected but not recorded (which translates into the non availability of the \emph{which-way} data to the experimenter/observer).
There are many possible set ups for this experiment. A simple instantiation would be to place (turned on) detectors  at the slits and turn off any  device recording the information sent from these detectors (or this could be simply done by unplugging cables transmitting impulses from the detectors to the recording device, the main idea for this experiment is to test the impact of ``detecting but not making the data available to an observer'').

This test could also be implemented with entangled pairs using the delayed choice quantum eraser experiment (see Figure \ref{figerasuresetup})
 by:
 \begin{itemize}
\item Simply removing the coincidence counter from the experimental setup and recording (only) the output of $D_0$ (result screen).  $D_0$ should display the wave pattern if the experiment is successful.
\item Or by turning off the coincidence counter channels $D_3$ and $D_4$ (and/or the detectors). If the experiment is successful, then $D_0$ should (without the available information for sorting/subsetting between $D_3$ and $D_4$) display an interference pattern (and sorting the impacts at $D_0$ by $D_1$ or $D_2$ should also show interference patterns).
\end{itemize}

\paragraph{On the  availability of the \emph{which-way} data.}
These tests are designed based on the  conjecture  that it is ``the availability of objective \emph{which-way} data at the time of observation of the interference/particle pattern''  that determines the nature (particle or wave) of the observed pattern. Note that ``availability of the \emph{which-way} data'' is implied by the \emph{which-way} data being recorded on objective media (but does not imply a simple ``observation of the which-way data''). This distinction must be used as an Occam's razor to avoid the history problem that plagued solipsism and Wigner's theory.
For instance if experimenter I watches only the pattern-screen and experimenter II watches the \emph{which-way}-screen, then even if the pattern-screen watcher does not know the \emph{which-way} data,  the pattern screen should show a wave pattern (since \emph{which-way}
data is only available as subjective information recorded in the memory of experimenter II).
In general, the notion of ``availability of the \emph{which-way} data'' must be separated from the notion of consciousness (as represented by one or more experimenters) ``knowing'' \emph{which-way} data by means of unrecorded observation. For instance if experimenter II is replaced by a USB flash drive, then experimenter I should see a particle pattern since at the time of observation of the pattern,  \emph{which-way} information is available (recorded in the USB ash drive). Note that the time of observation of (wave or particle) pattern is a determining factor. For instance, in the following subsection, experimenters I and II are replaced by USB flash drives I and II then when USB I should show (a) a wave pattern if USB II is irreversibly destroyed prior to reading USB I (b) a particle pattern if USB II is available for reading when USB I is read.

 \begin{figure}[h!]
	\begin{center}
			\includegraphics[width=1\textwidth]{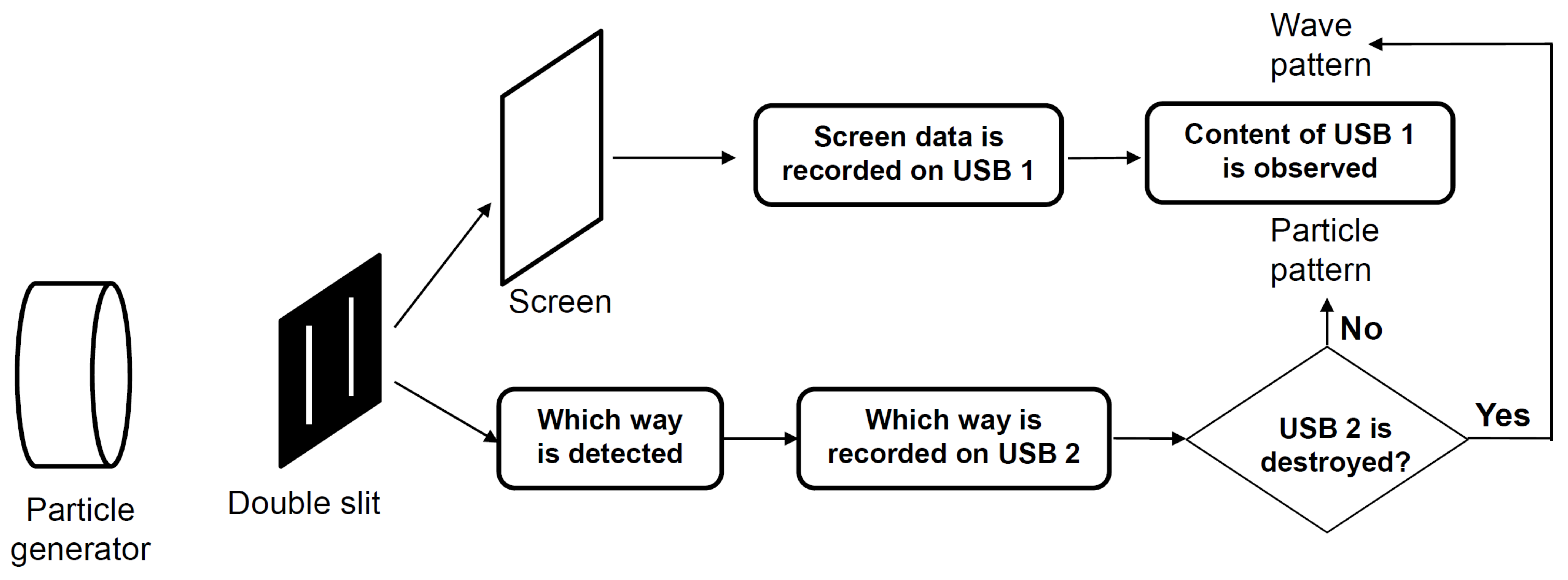}
		\caption{Erasing the \emph{which-way} data on a macroscopic scale}\label{figsub}
	\end{center}
\end{figure}

\subsection{Erasing the \emph{which-way} data on a macroscopic scale}\label{secpredict3}
\begin{figure}[h!]
	\begin{center}
			\includegraphics[width=1\textwidth]{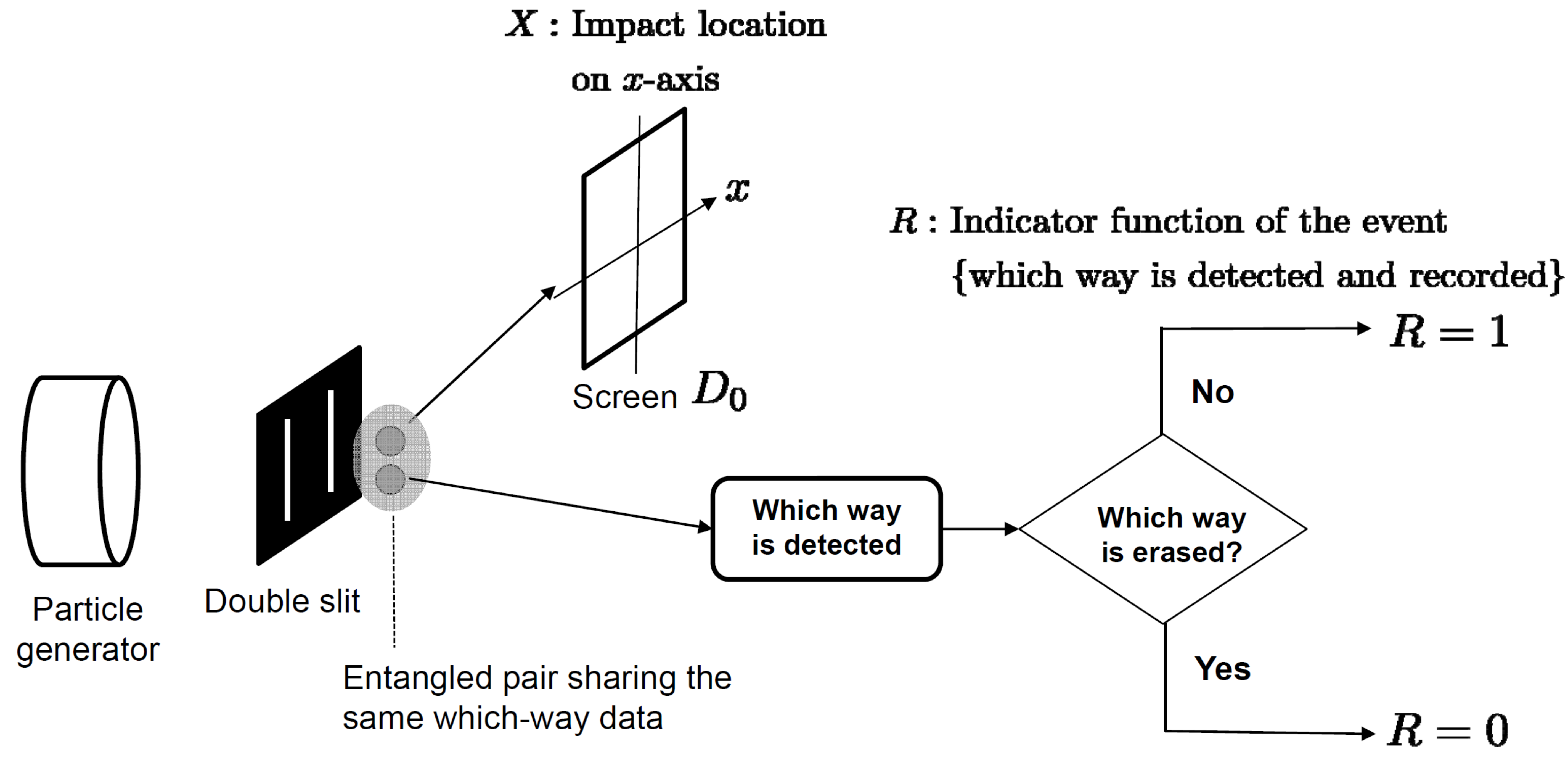}
		\caption{Delayed Erasure Experiment. \emph{Which-way} data is randomly recorded with probability $1/2$.}\label{figsuperdelayedchoicequantumeraser}
	\end{center}
\end{figure}

In the proposed experiment, illustrated in a simplified and conceptual form in Figure \ref{figsub}, the decision to erase the \emph{which-way} data is delayed to a macroscopic time-scale.
This can be implemented by using the classical double slit experiment shown in Figure \ref{figdoubleslit} where the recordings of the \emph{which-way} data and the screen data (impact pattern) are collected on two separate
USB flash drives.
By repeating this process $n$ times one obtains $n$ pairs of USB flash drives ($n$ is an arbitrary non-zero integer). For each pair, the  \emph{which-way} USB flash drive is destroyed with probability $p_d=1/2$. Destruction must be such that the data is not recoverable and no trace of the data is left on the computer that held and transferred the data. For $n$ even, one can replace the coin flipping randomization by that of randomly selecting a subset composed of half of the pairs of USB flash drives containing \emph{which-way} data for destruction (with uniform probability over such subsets). The test is successful if the USB flash drives storing impact patterns show an interference pattern only when the corresponding \emph{which-way} data USB flash drive has been destroyed.
This test can also be performed by using the delayed choice quantum eraser experiment or its modified version illustrated in Figure \ref{figsuperdelayedchoicequantumeraser}.  For this implementation, one USB flash drive is used to record the data generated by the photons for which $X$ is measured (output of $D_0$) and other USB flash drives to record the data generated by  $D_1, D_2, D_3$ and $D_4$ along with the associated output of the coincidence counter.

\subsection{Predicting erasure in the delayed choice quantum eraser experiment}\label{secpredict}
The proposed test is based on  a modification of the delayed choice quantum eraser experiment \cite{kim2000delayed}.
In this modification,  we use the facts that (1) the entangled pair of photons discussed in  Remark \ref{rmkfeature} share the same \emph{which-way} data (2) the experiment can be arranged so that the  first photon  hits the screen (sending a pulse toward the coincidence counter) before the second one reaches the beam-splitter causing the erasure or recording of the \emph{which-way} data with probability $1/2$ (but the time interval between these two events must be significantly smaller than the time interval between creation of photon pairs to preserve the information provided by the coincidence counter).
The location $X$ of the impact (on the $x$-axis) of the first photon on the screen (see Figure \ref{figsuperdelayedchoicequantumeraser}) is then recorded and used to predict whether the \emph{which-way} information will be erased ($R=0$) or kept/recorded ($R=1$).
More precisely by applying Bayes' rule we obtain that
 $\P[R=1|x\leq X\leq x+\delta x]=\frac{\P[R=1]}{\P[x\leq X\leq x+\delta x]} \P[x\leq X\leq x+\delta x|R=1]$.
Using $\P[x\leq X\leq x+\delta x]=\P[x\leq X\leq x+\delta x|R=0] \P[R=0]+
\P[x\leq X\leq x+\delta x|R=1] \P[R=1]$ and $\P[R=0]=\frac{1}{2}$ we deduce that
\begin{equation}
\P[R=1|x\leq X\leq x+\delta x]= \frac{1}{1 + f(x)} \text{ with }f(x)=\frac{\P[x\leq X\leq x+\delta x|R=0]}{\P[x\leq X\leq x+\delta x|R=1] }.
\end{equation}
Let $d$ be the distance between the two slits and $L$ the distance between the slits and the screen (where $X$ is recorded). Write $\lambda$ the wavelength of the photons and $a := \frac{\lambda L}{d}$.
Using the standard approximations $\P[x\leq X\leq x+\delta x|R=1]\approx 2 I_0 \,\delta x$ and
$\P[x\leq X\leq x+\delta x|R=0]\approx 4 I_0 \cos^2(\pi \frac{x}{a})\,\delta x$ (valid  for $x\ll L$) we obtain that
\begin{equation}
\P[R=1|x\leq X\leq x+\delta x]\approx  \frac{1}{1 + 2 \cos^2(\pi \frac{x}{a})}\,.
\end{equation}
Therefore if the proposed experiment is successful, then the distribution of the random variable $R$ would be biased by that of $X$ and this bias could be used by a microprocessor whose output would predict the value of the random variable $R$ (prior to its realization) upon observation of the value of $X$.
This bias is such that, if the value of $X$ corresponds to a dark fringe of the interference pattern and a high intensity part of the particle pattern, i.e. if $\cos(\pi \frac{x}{a}) \approx 0$ and $x/a\approx 0$, then the photon must be reflected at  $BS_a$ and $BS_b$ (i.e. $R=1$) with a probability close to one.
Observe that the value of $R$ is determined by whether the photon is reflected rather than  transmitted the beam splitters $BS_a$ and $BS_b$ (which are large masses of materials that could be at large distance from the  screen $D_0$).
Therefore, if the proposed experiment is successful, then for values of $X$ corresponding to a dark fringe of the interference pattern,
it would appear as if the measuring, recording, and observing of impact location $X$ determines whether the \emph{which-way} data will or will not be erased. Such a result would solve the causal flow of time issue in delayed erasure experiments: detection at $D_0$ would now determine (or introduce a bias in) the choice, i.e. reflection or  transmission, at $BS_a$ and $BS_b$.  However, a new issue would be created: The detection at $D_0$ deterministically selecting (or, for a general value of $X$, strongly biasing the probability of) the choice at $BS_a$ and $BS_b$ (reflection or transmission) when that choice is supposed to be random (or, for a general value of $X$, independent from $X$).   Although this could be seen as a paradox such a result would have a very simple explanation in a ``simulated universe'': the values of $X$ and $R$ are realized  at the moment the recorded data becomes available to the observer (experimenter).

 \begin{figure}[h!]
	\begin{center}
			\includegraphics[width=1\textwidth]{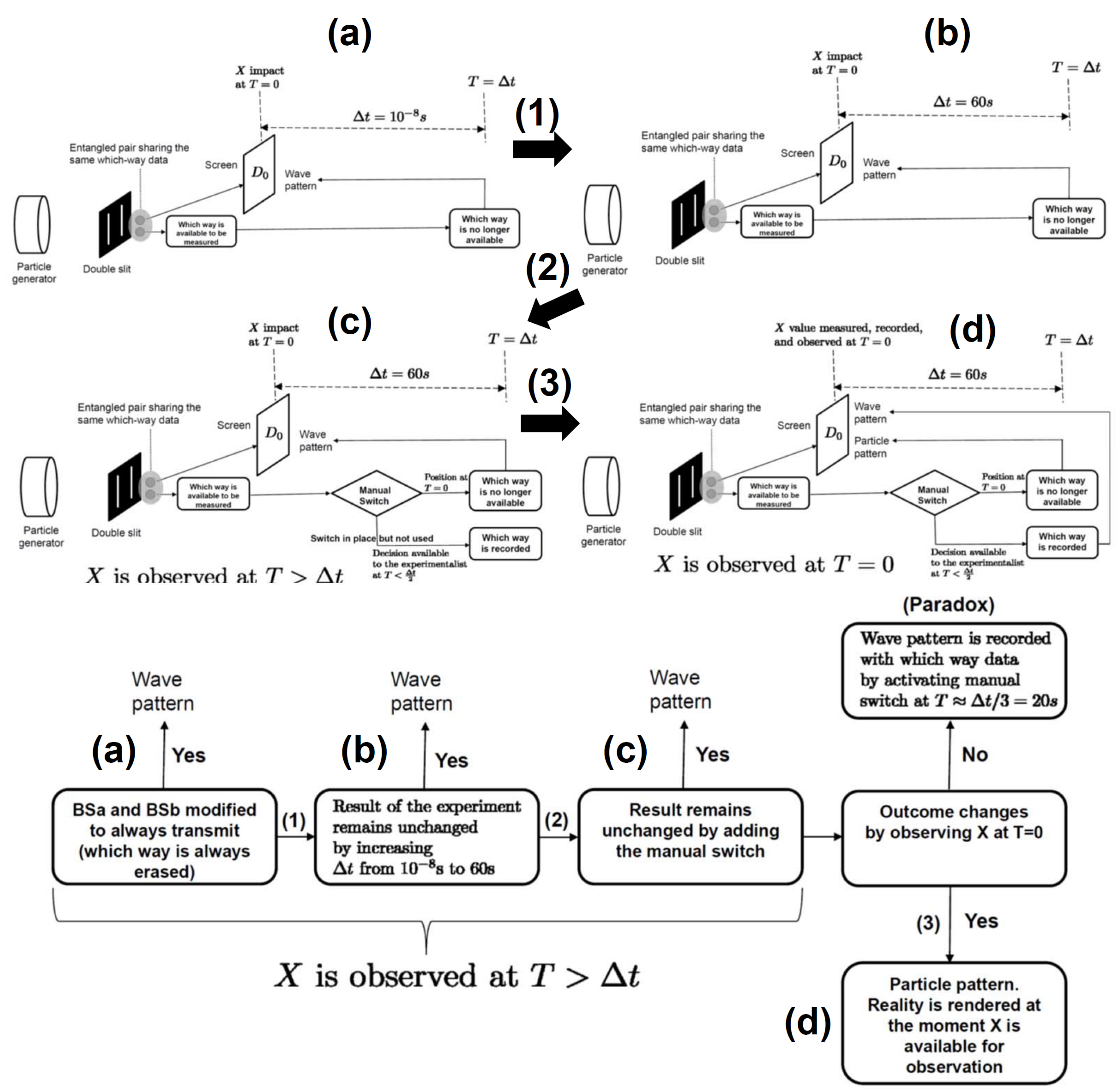}
		\caption{Testing the role of the observer}\label{figparadox}
	\end{center}
\end{figure}

\subsection{Exploiting conflicting requirements of consistency preservation and detection avoidance}\label{seclkjhjhu}
The purpose of the thought experiment (illustrated in a simplified and conceptual form in Figure \ref{figparadox}) described here is not only to
test the role of the observer in the outcome of a variant of the delayed choice quantum eraser experiment, but also to show that if the conscious observer/experimenter plays no role in the outcome then  the rendering of reality would have significant discontinuities.
More precisely, we will use the logical flow of this thought experiment to prove, per absurdum, that at least one of the following outcomes must hold true.
\begin{enumerate}[(I)]
\item  Steps (1) or (2) in Figure \ref{figparadox} do not hold true (which would be a discontinuity in the rendering reality).
\item Reality is rendered at the moment the corresponding information becomes available for observation by an experimenter (which would be an indication that the simulation (VR) theory is true).
\item \emph{Which-way} data can be recorded with a wave pattern (which would be a paradox).
\end{enumerate}

Consider  the delayed choice quantum eraser experiment  \cite{scully1982quantum, kim2000delayed} illustrated in Figure \ref{figdelayedchoicequantumeraser}.  Let $\Delta t$ be the interval of time (in the reference of lab where the experiment is performed) separating the impact of the first (signal) photon on the screen $D_0$ from the  moment the second (idler) photon reaches the beam splitter $BS_c$ causing the \emph{which-way} data to be either available (recorded) or not available (erased).
The experiments of X. Ma, J. Kofler, A. Qarry et al. \cite{ma2013quantum} suggest that the set-up could be such that $\Delta t$ could be arbitrarily large without changing the outcome of the delayed erasure (we will make that assumption). We will also assume that the time interval between the production of entangled pairs of photons can be controlled so that during each time interval $\Delta t$, only one pair of photons runs through the experiment.

Write $\P_\wa$ the probability distribution on $X$ associated with a wave pattern.  Write $\P_\pa$ the probability distribution on $X$ associated with a particle pattern.
For $I$ a subset of the possible values of $X$, write
\begin{equation}\label{eqkjhjhhkj}
\delta(I):=\P_\pa[X\in I]+ \P_\wa[X\not \in I]
\end{equation}
Assume that the distance separating the two slits and the distance separating the screen $D_0$ from the two slits are such that $I$ can be chosen so that $\delta(I)<0.9$. Observe that $\delta(I)=1+\P_\pa[X\in I]- \P_\wa[X \in I]$, therefore $\min_{I}\delta(I)=\operatorname{TV}(\P_\pa,\P_\wa)$ where $\operatorname{TV}(\P_\pa,\P_\wa)$ is the total variation distance between $\P_\pa$, and $\P_\wa$. Therefore  the possibility of choosing $I$ so that $\delta(I)<0.9$ is equivalent to $\operatorname{TV}(\P_\pa,\P_\wa)>0.1$

\begin{enumerate}[(a)]
\item Remove the beam splitters $BS_a$ and $BS_b$ (or modify them to be totally transparent) so that \emph{which-way} data is not available. The experimentalist observes the outcome of the experiment after X has been realized and the possibility of which-way data has been eliminated. One should get an interference pattern at $D_0$, as illustrated in Figure \ref{figparadox}-(a).
\item Increase $\Delta t$ from $\Delta t \approx 10^{-8}$s to $\Delta t \approx 60$s. The totally transparent beam splitters $BS_a$ and $BS_b$ occur on the timeline at $\Delta t/2$.  The experimentalist observes the outcome of the experiment after $X$ has been realized and the possibility of \emph{which-way} has been eliminated. If the outcome of the experiment does not depend on $\Delta t$ then one should get an interference pattern at $D_0$, precisely as it did in \ref{figparadox}-(a) and as illustrated in Figure \ref{figparadox}-(b).

\item
Introduce, as illustrated in Figure \ref{figparadox}-(c), a manual switch that, when activated, causes beam splitters $BS_a$ and $BS_b$ to totally reflect (become mirrors) so that \emph{which-way} data will always be collected and remain available.  This manual switch gives the experimentalist the option to produce and record (or not) \emph{which-way} data by activating the switch (or not) at $T < \Delta t/2$).
Assume the switch can be quickly activated or not at the (arbitrary) decision of the experimentalist.
If the position of the switch at $T=0$ leads to erasure of the \emph{which-way} data,  and if the experimentalist observes the outcome of the experiment at $T>\Delta t$ (after $X$ has been realized and \emph{which-way} has been erased), then  an interference pattern should be observed as illustrated in Figure \ref{figparadox}-(c).

\item If the experimentalist observes the value of $X$ at $T=0$ instead of at $T>\Delta t$. Then the following outcomes are possible
\begin{enumerate}[i]
\item  $X$ is sampled from $\P_\wa$ (an interference pattern) when the switch is inactive and from $\P_\pa$  (a particle pattern) when the switch is active.
\item $X$ is always sampled from $\P_\pa$ (a particle pattern).
\item $X$ is always sampled from $\P_\wa$ (an interference pattern).
\item Not (i), (ii) or (iii).
\end{enumerate}
Assume that alternative (i) holds.  Let the experimentalist implement the following switch activation strategy with $I$ chosen so that $\delta(I)<0.9$.
\begin{Strategy}\label{str1}
Use the following algorithm
\begin{itemize}
\item If $X\not \in I$ then do not activate the switch (let \emph{which-way} be erased).
\item If $X \in I$ then  activate the switch (record \emph{which-way}).
\end{itemize}
\end{Strategy}
Let $\P$ be the probability distribution of $X$ in outcome (i). Observe that
\begin{equation}\label{eqkhjhkjkjhui}
1=\P[X\in I]+\P[X\not \in I]=\P_\pa[X\in I]+ \P_\wa[X\not \in I]=\delta(I),
\end{equation}
which is a contradiction with $\delta(I)<0.9$, and therefore outcome (i) cannot hold. Outcome
(iv) would be a discontinuity. Outcome (iii) would allow the experimentalist to always activate the switch and record \emph{which-way} with an interference pattern as illustrated in Figure \ref{figparadox}-(d). Outcome (ii) would would be a strong indicator that this reality is simulated. Indeed if  one gets a particle pattern at
 $D_0$ independent of the position of the switch, then the observation at $T=0$ would have been the cause since this  would be the only difference between (c) and (d) if the switch is not activated.
\end{enumerate}
If the outcome of the experiment of Figure \ref{figparadox}-(c) is a wave pattern and that of  Figure \ref{figparadox}-(d) is a  particle pattern then the test is successful: the outcome is not entirely determined by the experimental/detection set-up and $X$  (reality/content) must be realized/rendered at the moment when \emph{which-way} becomes available to an experimenter/observer.
This experiment is likely to be successful in the sense that the  only  possible outcomes are: the exposure of  discontinuities in the rendering of reality, or paradoxes.

\paragraph{Clarification of the notion of pattern in Figure \ref{figparadox}.}
Since in the experiments illustrated in Figure \ref{figparadox}, samples/realizations $X_i$ of $X$ are observed one ($X_i$/photon) at a time (at $T=0$ or for $T>\Delta t$), we define  ``pattern'' as the pattern formed by a large number $n$ of samples/realizations $X_1,\ldots,X_n$ of $X$.
In Figures \ref{figparadox}-(a), \ref{figparadox}-(b) and \ref{figparadox}-(c) these samples are observed after the erasure of the \emph{which-way} data and the resulting aggregated pattern (formed by $X_1,\ldots,X_n$ for large $n$) must be that of an interference pattern.
In the experiment illustrated in Figure \ref{figparadox}-(d) samples/realizations of $X$ are observed one at a time, and the experimenter can decide after observing each  $X_i$ to record (by turning the switch on) the corresponding \emph{which-way} data or let that information be erased (by leaving the switch in its initial off state). Since the experimenter can base his decision to activate the switch at any step $i$ on the values of $X_1,\ldots,X_i$, the experimenter can implement an activation strategy such that the pattern formed by the subset of elements of $\{X_1,\ldots,X_n\}$ with switch \emph{on} is, to some degree, arbitrary (e.g. create a 3 slit pattern by activating the switch only when the value of $X$ is in 3 predetermined narrow intervals). Similarly the experimenter can implement an activation strategy such that the pattern formed by the subset of elements of $\{X_1,\ldots,X_n\}$ with switch \emph{off} is, to some degree, arbitrary. However he has no control over the pattern formed by all the elements $\{X_1,\ldots,X_n\}$ (with switch positions \emph{on} or \emph{off}).
Either
\begin{enumerate}
\item The pattern formed by aggregates of the values of $X_1,\ldots,X_n$ is independent of the positions of the switch (at all steps $1,\ldots,n$), if each $X_i$ is observed at $T=0$, and is that of a particle pattern (due to the availability of the which-way information to the experimenter at T=0). In particular, in the experiment of  Figure \ref{figparadox}-(d), the experimenter may always keep the switch \emph{off} so that none of the samples has a paired/recorded \emph{which-way} data and he would still obtain a particle pattern. This is not a paradox since the rendering is triggered through the availability of \emph{which-way} at $T=0$.
There is also no contraction with the suggested outcome of the experiment of Figure \ref{figrecordoff} since in
that experiment  the recording of the \emph{which-way} data is determined prior to the realization of $X$.
\item Or the pattern formed by aggregates of the values of $X_1,\ldots,X_n$ depends on the positions of the switch (at all steps $1,\ldots,n$) that are, at each step $i$, determined by the experimentalist $20s$ after the observation of $X_i$ (i.e. not produce an $X_i$ in a dark fringe of the diffraction pattern if the experimentalist decides $20s$ later to not activate the switch, which would be the paradoxes discussed around Strategy \ref{str1} since the activation strategy is arbitrary).
\end{enumerate}

\paragraph{Difference between the experiment of Figure \ref{figsuperdelayedchoicequantumeraser} and that of Figure \ref{figparadox}-(d).}
Observe that in the experiment illustrated in
Figure \ref{figsuperdelayedchoicequantumeraser}, the value of $X$ is used at $T=0$ (by a microprocessor, $\mu$-p) to predict the later value of $R$ (i.e., the erasure or recording of \emph{which-way}). In Figure \ref{figparadox}-(d), the value of $X$ is observed by an experimenter before deciding whether \emph{which-way} should be erased or recorded. Although in both experiments the value of $X$ seems to be \emph{operated on} at $T=0$  two different outcomes should be expected if the simulation theory is true based on  the analysis of how a VR engine would operate. In
Figure \ref{figsuperdelayedchoicequantumeraser} the pattern at $D_0$ formed by the subset of elements of $\{X_1,\ldots,X_n\}$ for which $R=0$ is that of an interference pattern; and the pattern at $D_0$ formed by the subset of elements of $\{X_1,\ldots,X_n\}$ for which $R=1$ is that of a particle pattern. In Figure \ref{figparadox}-(d) the pattern at $D_0$ is always that of a particle pattern (independently from the decision of the experimenter to activate the switch and record the data). This difference is based on the understanding that, if the decisions of the experimenters are external to the simulation, then, while in the experiment of Figure \ref{figsuperdelayedchoicequantumeraser} the VR engine would be able to render the values of $X$ and $R$ at the same moment to the experimenter (since the microprocessor using the value of $X$ would be part of the simulation), the VR engine would not necessarily be able to predict the (arbitrary) decision (that may or may not depend on the value of $X$) of the experimenter (to activate the switch) in the experiment proposed in Figure \ref{figparadox}-(d) (\emph{which-way} is available for observation by the experimenter at $T=0$). This difference could also be understood as a clarification of the notion of availability of \emph{which-way} data in a VR.

\paragraph{Control of the switch by a microprocessor.} The switch of Figure \ref{figparadox}-(d) could in principle be activated by microprocessor. In that setup the time interval $\Delta t$ could be significant reduced from the value proposed in  Figure \ref{figparadox}-(d). Since Strategy \ref{str1} would still be available for implementation (as an algorithm), Alternative (i) discussed in the outcome of Figure \ref{figparadox}-(d) would still lead to a contradiction. Alternative (iii) would allow us to record \emph{which-way} with an interference pattern. Alternative (iv) would be a discontinuity.
Alternatives (ii) or (iii) would an indicator of an intelligent VR engine reacting
to the intention of the experimentalist. Alternative (i) leads to a logical paradox for $\delta(I)<0.9$.

Note that Eqn.~\eqref{eqkhjhkjkjhui} would still be a contradiction if $\delta(I)<1$ (and the existence of such an $I$ is ensured by $\operatorname{TV}(\P_\pa,\P_\wa)>0$). We use $\delta(I)<0.9$ to account for experimental noise. Although we cannot predict the outcome of the proposed experiment, we can prove based on Eqn.~\eqref{eqkhjhkjkjhui} that the pattern produced at the screen $D_0$ cannot be the result of sampling $X$ from a particle distribution when the switch is active and a wave distribution when the switch is inactive. Therefore, although the experiment has not been performed yet we can already predict that its outcome will be new. One possible outcome is that the $X$ will be sampled from a particle distribution independently of the position of the switch which would also be an indicator of a VR engine reacting to the intent of the experiment.

\subsection{Further questions and further tests}
The simplest test for of the simulation theory is the experiment proposed in Subsection \ref{secpredict2} where \emph{which-way} is detected but not made available for observation or recording.
 Although a positive outcome for the experiment of Figure \ref{figrecordoff} would support the validity of the simulation theory,  it would also call for more questions and more tests to characterize the process of rendering of reality.
Some of these questions are listed below,
\begin{enumerate}
\item For \emph{which-way} datum  to cause its corresponding dot on the pattern screen to be in a particle pattern, does it have to be correlatable (usually in time) with its associated  point on the  pattern screen, or is only a recording the (\emph{which-way}) datum’s existence sufficient?
\item	Does anonymous (unlabeled) \emph{which-way} data cause its corresponding dot on the pattern screen to be in a particle pattern?
\item	Is the objective recording of \emph{which-way} information (e.g on a hard drive) necessary to  produce a particle pattern, or is subjective observation (e.g. in the imperfect memory of an experimenter) of the \emph{which-way} data sufficient?
\end{enumerate}

By using the same strategy as in subsection \ref{seclkjhjhu} it is possible to show that if there is a difference between the objective and subjective recording of \emph{which-way} (i.e. objective recording leads to a particle pattern and subjective recording leads to a wave pattern) then the VR server would have to adjust its rendering to the intent of the experiment to avoid creating a paradox
(which would indicate that the VR server, the source of the VR,  is conscious).
The proof is as follows. Assume that \emph{which-way} data is collected only on a perishable media that persists for $\Delta T$ seconds (after which the data is lost permanently). Assume that during this interval of time the experimenter has the option to record the data permanently (on a hard drive).
Assume for the sake of the clarity of argument that the impact locations of the wave pattern and the particle pattern do not overlap (the argument can be extended to the overlapping case using  probabilistic inequalities as in subsection \ref{seclkjhjhu}). More precisely assume that there exists a portion $I$ of the pattern screen such that if the impact is in $I$ then it must be part of a wave pattern and if it is not in $I$ then it must be part of a particle pattern.
Now for each impact let the experimenter  follow the rule: (1)
If the impact is in $I$ then objectively record \emph{which-way}  data (on a hard drive)
(2) If the impact is not in $I$ then do not objectively record \emph{which-way}  data (simply observe it).
Then by  definition of the set $I$, the pattern formed by impacts whose \emph{which-way} data has not been objectively recorded cannot be a wave pattern, which implies that either (a) there is no difference between objective and subjective recording (both lead to a particle pattern and no impact is observed in $I$) (b) the VR server can adjust its rendering to the intent of the experiment.
Note also that if the proposed rule (record  \emph{which-way} if and only if the pattern impact is in $I$) is  enforced by an algorithm (i.e. if the experimenter can be taken out of the loop), then the resulting paradox seems to reinforce the idea that the VR server would have to adjust its rendering based on the intent of the experiment (since as suggested by the quantum eraser experiment and discussed in Subsection \ref{seclkjhjhu}, detection of \emph{which-way}, seems to not be sufficient to ensure a particle pattern).

\paragraph{Acknowledgments.} We thank Lorena Buitrago for her help with Figure \ref{figerasuresetup}. We also thank an anonymous referee whose detailed comments and suggestions have lead to significant improvements.

\bibliographystyle{plain}
\bibliography{merged,RPS}

\end{document}